\documentclass[conference]{IEEEtran}
\IEEEoverridecommandlockouts
\usepackage[english]{babel}
\usepackage{graphics, graphicx}
\usepackage{amsmath, amsthm, amssymb, amsfonts}
\usepackage{algorithm}
\usepackage{algpseudocode}
\usepackage{multirow}
\usepackage{url}
\usepackage{soul}  
\usepackage{tabularx}
\usepackage{caption}
\usepackage{subcaption}
\usepackage{comment}
\usepackage{tikz}
\usepackage{cite}
\usepackage{xcolor}
\usepackage{tabu}
\usepackage{tabularx}
\usepackage[nocomma]{optidef}
\usepackage{soul}

\usepackage{pdfpages}
\usepackage{enumitem}
\definecolor{gold}{rgb}{0.85,.66,0}
\newcommand{\ta}{\textcolor{blue}}
 \newcommand{\hls}{\textcolor{black}}

\newcommand{\bPh}{\boldsymbol{\Phi}}

\DeclareMathOperator*{\argmax}{arg\,\max}

\def\BibTeX{{\rm B\kern-.05em{\sc i\kern-.025em b}\kern-.08em T\kern-.1667em\lower.7ex\hbox{E}\kern-.125emX}}

\title{LSTM-ACB-Based RA for IoT Mixed Traffic}

 \author{

\IEEEauthorblockN{{Herman Lucas dos Santos}}
\IEEEauthorblockA{\textit{Electrical Eng. Dept.}
\textit{UEL}\\
Londrina, Brazil.\\
hermanlds@gmail.com
\vspace{-6mm}}

\and

\IEEEauthorblockN{{Cristiano Magalh\~aes Panazio}}
\IEEEauthorblockA{\textit{Dept. of Telecom. and Control}
\textit{EP-USP}\\
S\~ao Paulo, Brazil.\\
cpanazio@usp.br
\vspace{-6mm}}
\and

\IEEEauthorblockN{Jos\'e Carlos Marinello Filho}
\IEEEauthorblockA{\textit{Electrical Eng. Dept.}
\textit{UTFPR}\\
Corn\'elio Proc\'opio, Brazil.\\
jcmarinello@utfpr.edu.br\vspace{-6mm} }

\and

\IEEEauthorblockN{Taufik Abr\~ao}
\IEEEauthorblockA{\textit{Electrical Eng. Dept.}
\textit{UEL}\\
Londrina, Brazil.\\
taufik@uel.br
\vspace{-6mm}}

 \thanks{This work was supported by the National Council for Scientific and Technological Development (CNPq) of Brazil under Grants 405301/2021-9, 141485/2020-5, and 310681/2019-7.}
 }

\begin{document}

\maketitle

\begin{abstract}
A novel random access (RA) scheme for mixed URLLC-mMTC traffic scenario is proposed using realistic statistical models, with the use mode presenting long-term traffic regularity. The traffic is predicted by a long short-term memory neural network, which enables a traffic-aware resource slicing aided by contention access class barring-based procedure. The method combines a grant-free (GF) RA scheme with an intermediate step to congestion alleviation. The protocol trade-off is a small overhead \hls{while enabling} a higher number of decoded received packets thanks to the intermediate step. Numerical results \hls{evaluate} the system performance for each procedure and combined solution. A comparison with GF benchmark reveals substantial \hls{improvement} in system performance.
\end{abstract}

\begin{IEEEkeywords}
Random access protocols, IoT networks, network slicing, long-short term memory (LSTM).
\end{IEEEkeywords}
\vspace{-3mm}
\section{Introduction}
\vspace{-1mm}
{Mixed traffic in fifth-generation (5G) telecommunication systems are challenging because of distinctive service requirements, as in mMTC and URLLC co-existence \cite{Petar2018}. {Resource granting for different use modes, \hls{namely resource slicing (RS)}, is challenging due to their individual restriction}. Statistical models for the services were studied in \cite{Thota2019,Alsenwi2021}. \hls{RS manages \textit{time} (T) and \textit{frequency} (F) resources} to support different services. \hls{URLLC use mode} demands fast and reliable transmission, requiring high bandwidth, and \hls{mMTC use mode} a huge number of resource blocks (RBs) for high connectivity. 3GPP release 15 enabled the use of RS schemes with \textit{flexible numerology}, \textit{i.e.} using sub-carrier spacing (SCS) higher than 15 kHz \cite{Memisoglu2021}. Under mixed traffic, RS efficiently \hls{allocates} T-F resources \cite{Alsenwi2021,Tun2020}. }

{Furthermore, access control optimization (ACO) \hls{exploits} the network \textit{backlog}, \textit{i.e.} number of users attempting to access the network, to improve traffic control and resource utilization \cite{Sim2020}. These procedures are either: a) machine learning (ML)-based \hls{or} b) non-ML-based. A classic ACO scheme is the access class barring (ACB), which \hls{uses} statistics \hls{to control the user equipment (UE) access attempts efficiently.}  }

{This paper proposes \hls{an} RA protocol assisted by traffic predictor, RS, and ACB-based contention alleviation for mixed mMTC-URLLC 5G use modes. We employ \hls{an} LSTM in traffic prediction, \hls{a bin-packing based heuristic supporting different packet size and latency requirements, and a novel three-step RA protocol with congestion alleviation procedure.}}
\section{System Model}
\label{sec:SystemModel}

{Consider a time-division duplex M-MIMO system constituted of mixed URLLC-mMTC traffic composed by a cell containing a BS which serve $\mathcal{K} = \{K^{\text{m}},K^{\text{u}} \}$ mMTC and URLLC UEs, respectively, where the mMTC have a \hls{massive} number of devices with low activation probability and URLLC follows a \hls{periodic} distribution.}

\vspace{1mm}

{\noindent\textbf{\textit{System Traffic}}: Each service \hls{has} $P^m,P^u$ packet size, in bytes, for mMTC and URLLC, respectively, and $P^m>P^u$ \cite{Alsenwi2021}. We model mMTC devices arrival as a uniform variable with activation probability $p$ plus a set of $K_m^p$ devices generating one packet every $T_m$ frames \cite{Jiang2019}, and URLLC traffic as a repeating Beta distribution in $T_u$ frames. Both arrivals are modeled as:}
\vspace{-2mm} 
\begin{multline}
{\dot{K}^m} =  \sum\limits_{i = 1}^{K^{\text{m}}-K^p_m} \left[\mathcal{U}[0,1) \geq p\right]_i + K_m^p \sum\limits_{k=-\infty}^\infty \delta (t-kT_m),
\label{eq:mMTCTrafficModel} 
\end{multline}
\vspace{-4mm}
\begin{multline}
{\dot{K}^u} = \sum\limits_{i = 1}^{K^{\text{u}}} \left[\mathcal{U}[0,1) \geq 1- \frac{t^{\alpha-1}(T_u-t)^{\beta-1}}{T_u^{\alpha+\beta-1}\text{Beta}(\alpha,\beta)}\right]_i,
\label{eq:URLLCTrafficModel}
\end{multline}
{\noindent where $\delta$ is \hls{Dirac} impulse function, $\text{Beta}(\alpha,\beta)$ is the Beta function with $\alpha$ and $\beta$ parameters. The total number of active devices \hls{including} new arrivals and re-transmissions from previous frames unsuccessful access attempts, \textit{i.e.} \textit{network backlog,} is represented $\breve{K}^{\text{m}}$ and $\breve{K}^{\text{u}}$ for each use mode. }

\vspace{1mm}
{\noindent\textbf{\textit{Resources Slicing}} 5G New Radio presented larger \textit{sub-carrier spacing} (SCS) than LTE 15 kHz as the numerology, where $\Delta f = 2^\mu \times 15$ kHz, with $\mu = \{0, 1, \dots, 4\}$ the numerology factor, which can reduce {\it transmission time interval} (TTI) and/or enhance the quantity of information transmitted under a given fixed time frame once more bandwidth is used. A time frame $t$ \hls{has} the duration of 10 ms divided \hls{into} $S$ time-slots and $F$ frequency sub-carriers, and a single 15 kHz, 1 ms resource block (RB) carries $\nu = 14$ OFDM symbols. The TTI, in symbols, and the quantity of information in one time slot is: }
$\text{TTI} = \frac{{N}_{\text{sym}}}{2^\mu \cdot \nu }\,\, \text{[ms]},\quad \text{or} \quad N_{\text{sym}} =  2^\mu \nu \,\,\,\,\text{[symbols/ms]}$

\begin{figure}[!htbp]
\vspace{-4mm}
\centering
\includegraphics[width=.8\linewidth]{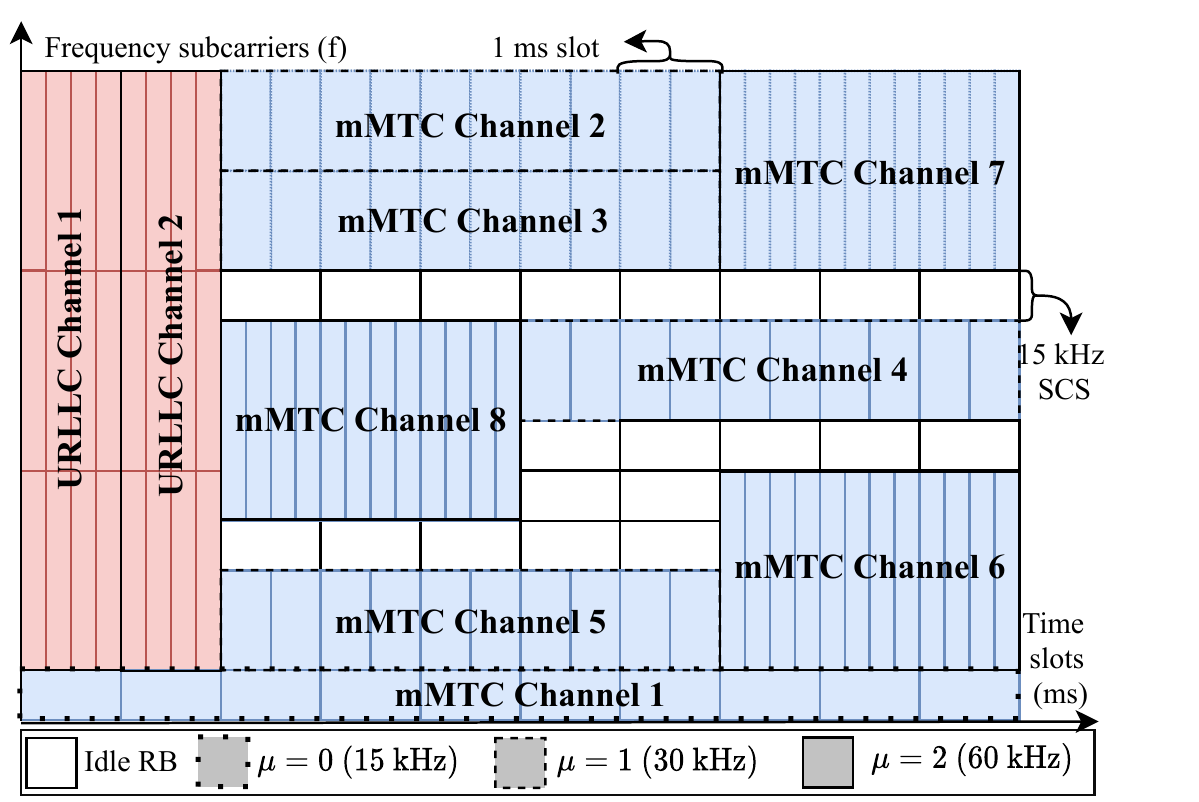}
\caption{\small {T-F RS in mixed mMTC-URLLC traffic with fixed TTI.} 
}
\label{fig:Numerology}
\end{figure}

{Using numerology, the BS is able to create channels with lower latency, creating higher numerology sub-channels and assigning them to channels, \textit{e.g.} Fig. \ref{fig:Numerology}, where the colors refer to the numerology used and URLLC channels 1 and 2 are composed by three $\mu=2$ sub-channels each, mMTC channel 1 use $\mu = 0$ (minimum SCS usage), mMTC channels 2 to 4 use $\mu = 1$ (typical). The 6th to 8th mMTC channels use $\mu = 2$ (granting faster transmission, but less efficient in terms of resource allocation). }

{We represent a single-user channel $l$ as the assignment of a T-F set, represented by {${\bPh_l^u}^{(F\times S)}$ and ${\bPh_l^m}^{(F\times S)}$} matrices, for URLLC and mMTC use mode\hls{s}, respectively, containing $\phi_{{f,s}}\in\{0,1\}$ elements indicating which {time-frequency slots  RBs} are assigned to the $l$th channel. The resources are dedicated to use modes, \textit{i.e} the users do not compete for them and the BS tries to accommodate all $\breve{K}^{\text{m}}+\breve{K}^{\text{u}}$ users in orthogonal channels. The set of channels $\mathcal{L}$ is the {disjoint} union of $\mathcal{L}^u$ and $\mathcal{L}^m$ sub-sets, and $|\mathcal{L}|=L=L^u+L^m$.}

\vspace{-2mm}

\section{Sub-Problem  Formulation and Solution}\label{sec:Problem}
\subsection{Backlog Prediction  {$(\mathcal{P}1)$}} \label{subsec:BacklogPred}
{The backlog is represented as $N^t=\breve{K}^{\text{m}} + \breve{K}^{\text{u}}$ active users in \hls{${t}$h} frame and the BS \hls{has} only the observation of previous frames channel states, defined by {\it i}) \textit{Success} when a single UE select a channel; {\it ii}) \textit{Collision} when more than one device select the same channel in a T-F frame; or {\it iii}) \textit{Idle}, when the channel \hls{was} not selected by any device. The channels states in a frame \hls{are} called \textit{observation}: }
 $O^{t} =  \big\{\{V_s^{u,t},V_c^{u,t},V_i^{u,t}\},\{ V_s^{m,t},V_c^{m,t},V_i^{m,t}\}\big\}$
{where $(u,m)$ indicate\hls{s} the use mode and $(s,c,i)$ refers to success, collision, and idle states in $t$th frame. $V$ is the number of channels with given states. The record of previous observations from $(t-T_\textsc{w})$th to $(t-1)$th frame, with $T_\textsc{w}$ being memory size, is:
$\mathcal{H}^t = \{O^{t-{{T_w}}}, O^{t-({T_w}-1)}, \dots, O^{t-1} \}$}
{The backlog of each class, \textit{i.e.} {\it number of active users}, {$\Breve{K}^u$ and $\Breve{K}^m$}, in a time frame can be estimated through a conditional probability problem that {consists} in finding the number with maximum probability of occurrence given the past observations. This problem is modeled as:}
 \vspace{-2mm} 
\begin{eqnarray} 
\hat{K}^{u,{(t)}}& = \underset{n^u \in \{0,1,\dots,K^u\}}{\argmax}  \mathbb{P} \{N^{u,{(t)}} = n^u|\mathcal{H}^{u,(t-1)}\}, \nonumber \\
\hat{K}^{m,{(t)}} & = \underset{n^m \in \{0,1,\dots,K^m\}}{\argmax}  \mathbb{P} \{N^{m,(t)} = n^m|\mathcal{H}^{m,(t-1)}\}, \nonumber\\
   \text{s.t.} & \hat{K}^{(t)}= \hat{K}^{u,(t)}+\hat{K}^{m,(t)}, \label{eq:backlogPred}
\end{eqnarray} 

\noindent consisting in a data driven problem, \textit{i.e.} estimated through observations and desired outputs.

\noindent\textbf{\textit{LSTM-aided Traffic Predictor Solution}} $(\mathcal{S}1)$. 
{LSTM is \hls{an} ML technique that can process the system history with low-complexity \cite{Jiang2019} and is suitable {since the traffic is statistically defined, \textit{i.e.} the statistics can be captured from observations. }
Given the three detectable states of channels, a {neural network} running LSTM architecture can be employed in such application. LSTM is composed by several layers containing a forget gate, input gate, output gate, and softmax output function.   }
\vspace{-1.5mm}
\subsection{Resource Slicing {$(\mathcal{P}2)$}} 
\vspace{-1mm}
{The RS is employed to {maximize} the \hls{number of} served users with orthogonal resources under URLLC UEs TTI and different packet sizes, with contiguous channels in T-F. We {model RS scheme} as a binary assignment problem, Eq. \eqref{eq:RBConfig},  $\omega^u > \omega^m \geq {\omega^p} \in[0; \, 1]$, with $\omega^u >\omega^m \geq \omega^p$, $\omega^u + \omega^m + \omega^p=1$ are priority scaling factors to URLLC and mMTC received packets and penalty for not fulfilling the demand for channels, respectively; $Z = \frac{F\cdot S - \iota^u\cdot K^u}{\iota^m}$, {where $\iota^i =  \left(\frac{P^i}{\log_2M^i}+\xi \right)\cdot \frac{1}{\nu}$ given modulation order $M^i$ and protocol overhead \hls{$\xi$}} is the package size in symbols for $i \in {u,m}$. {$\rho^u$ and $\rho^m$ are the number of assigned channels of each use mode, }
{$\mathbf{e}_1=\mathbf{1}^{1\times F},\, \mathbf{e}_2 = \mathbf{1}^{S\times 1}$}, and $\odot$ stands for the Hadamard product. ($\mathcal{P}2$) \hls{the} solution must optimize RB usage assuming known CSI in all T-F RBs given the priority of URLLC use mode and penalty for not assigning the maximum number of {channels 
given} the actual traffic $Z$, subject to the constraints;  the {elements of matrix $\bPh$ } in  constraint \eqref{eq:RBConfigc1} must be binary; in \eqref{eq:RBConfigc2}, each URLLC channel occupies only one time RB to guarantee the delay requirement and prevent resource wastage; \textit{\eqref{eq:RBConfigc3}} and \textit{\eqref{eq:RBConfigc4} the RBs in each sub-channel must spread in continuous time and/or frequency RBs} to ensure the transmission is not fragmented neither in T nor in F; \eqref{eq:RBConfigc5} \textit{the number of sub-bands must satisfy the $2^\mu$ numerology factor}; \textit{\eqref{eq:RBConfigc6} the assignment matrices should not overlap each other} (orthogonal resources); in \eqref{eq:RBConfigc7}, the sum of assigned RBs must satisfy the packet size requirement. }

\vspace{-5mm}
\begin{maxi!}|s|[2]{\bPh}{\sum\limits_{\ell=1}^{L^u} \omega^u \rho_{\ell}^u + \sum\limits_{\ell=1}^{L^m}\omega^m {\rho}_\ell^m {- \omega^p\left[\Breve{K}-\min\left(|L|,Z\right)\right]}} {\label{eq:RBConfig}}{}
\addConstraint{\phi_{f,s}}{\in \{0,1\}}{\label{eq:RBConfigc1}}
\addConstraint{\sum\limits_{i=1}^F \phi_{i,s}^u }{\leq 1}{\label{eq:RBConfigc2}}
\addConstraint{\Big\{\begin{matrix}
\phi_{k,s} = 1, \\
0,                      \end{matrix}}
{\begin{matrix}
\ {f_i \leq k \leq f_j} \\
 \text{ otherwise} \end{matrix}}{\label{eq:RBConfigc3}}
    \addConstraint{\Big\{\begin{matrix}
    \phi_{f,k} = 1, \\
 0,      \end{matrix}}{\begin{matrix}
 \ {s_i \leq k \leq s_j} \\
 \text{ otherwise} \end{matrix}}{\label{eq:RBConfigc4}}
\addConstraint{\mathbf{e}_1\boldsymbol{\phi}_{f}}{=k\cdot 2^\mu \ \forall \ \mu = \{0,1,2\}, \  k \in \mathbb{N}}{\label{eq:RBConfigc5}}
\addConstraint{\mathbf{e}_1\left[\bPh_i \odot \bPh_j\right]\mathbf{e}_2 = 0 }{ \ \forall \  i,j,\ i \neq j}{\label{eq:RBConfigc6}}
\addConstraint{\mathbf{e}_1\bPh^{\text{i}}\mathbf{e}_2}{\geq \iota^i, \  i \in \{u,m\}}{\label{eq:RBConfigc7}}
\end{maxi!}
\vspace{-5mm}

\noindent\textbf{\textit{RS based on Heuristic \textsc{MaxRect} $(\mathcal{S}2)$}}. The problem in \eqref{eq:RBConfig} can also be interpreted as a \textit{bin-packing} problem, once there is limited space, \textit{i.e.}, the T-F grid, and the channels can be packed as boxes with different sizes and should be arranged in a manner that the maximum of \textit{boxes} fit into this space \cite{Jukka}.
We adopt an heuristic approach based on the {\it maximal rectangles} {{\it bottom-left} (\textsc{MaxRect})} \cite{Jukka}, 
Algorithm \ref{alg:maxrect}.

\subsection{Congestion Control:  ACB $(\mathcal{P}3)$}

\hls{We propose a congestion alleviation tool based on ACB procedure after the UEs channel selection. In order to connect to the network, UEs perform an ACB check by generating a random number $q \sim \mathcal{U}[0,1)$ and  {comparing it to the ACB factor} $p_{\text{acb}}$; if $q\leq p_{\text{acb}}$ the device is permitted to continue in RA process, else it is shut down till next frame. }

\hls{
At each frame\ta{, 
$\breve{K}$} active UEs select one of the $L$ available channels. Assuming UEs have knowledge of their channel power, in channel selection they transmit with power control in a manner to compensate path-loss. Hence, the BS receives unitary power gain from each active UE. Hence, this power control scheme \ta{enable} 
the BS estimate perfectly how many UEs attempted to access the \ta{$\ell$-th 
channel}, \ta{$n^{(\ell)}$}. Thus, to maximize the number of transmissions, we derive an ACB factor as:
\begin{equation}
    p_{\rm acb}^{\ta{(\ell)}^*} = 1 - \frac{1}{n^{\ta{(\ell)}}}, \quad \ta{\ell=1,\ldots, L}
    \label{eq:opt_acb}
\end{equation}
}

\vspace{-6mm}
\begin{algorithm}[!htbp]
\small
\begin{algorithmic}[1]
\Require{$\hat{K}^u$, $\hat{K}^m$, $F$, $S$, $P^u$, $P^m$}
\Ensure{$\bPh$}
 \While{$\ell \leq \hat{K}^u$}
\State Find in subspace $\mathcal{R}$ the bottom-left vertex\;
\State Assign $\big( \frac{P^u}{\log_2 M^u} + \xi \big) \cdot \frac{1}{\nu} $ freq. in 1-time-slot to $l$th channel\;
\State $\ell = \ell+1$
\State $\mathcal{R} = \mathcal{R} - l$ \EndWhile
 \While{$[\ell \leq (\hat{K}^u + \hat{K}^m)] \text{ {\bf AND} } [{|\mathcal{R}|} \geq \big( \frac{P^m}{\log_2 M^m} + \xi \big) \cdot \frac{1}{\nu}]$}
\State Find in subspace $\mathcal{R}$ the bottom left vertex\;
\State Set $\mu$ = 0 and shape the channel \textit{box}\;
\State Assign the selected resources to the $l$th channels\;
\While{$l$th channel do not fit the resource grid}
\State $\mu = \mu+1$; recalculate the box\;
\If{$\mu = 2$}
\State Set $\mu = 0$; search for next bottom-left vertex in $\mathcal{R}$\;
        \EndIf
    \State $\ell = \ell+1$
    \State $\mathcal{R} = \mathcal{R} - \ell$ \EndWhile
    \EndWhile
 \caption{\textbf{\textsc{maxrect}} - Maximal Rectangles Algorithm}
 \label{alg:maxrect}
\end{algorithmic}
\end{algorithm}
\vspace{-6mm}
\subsection{{Full-Solution: LSTM-ACB-based RA Scheme with Slicing}}
\label{sec:RAScheme}

\begin{figure}[htbp!]
\vspace{-3mm}
\centering
\includegraphics[trim=1.8mm 0mm 20mm 0mm, clip,width=.8\linewidth]{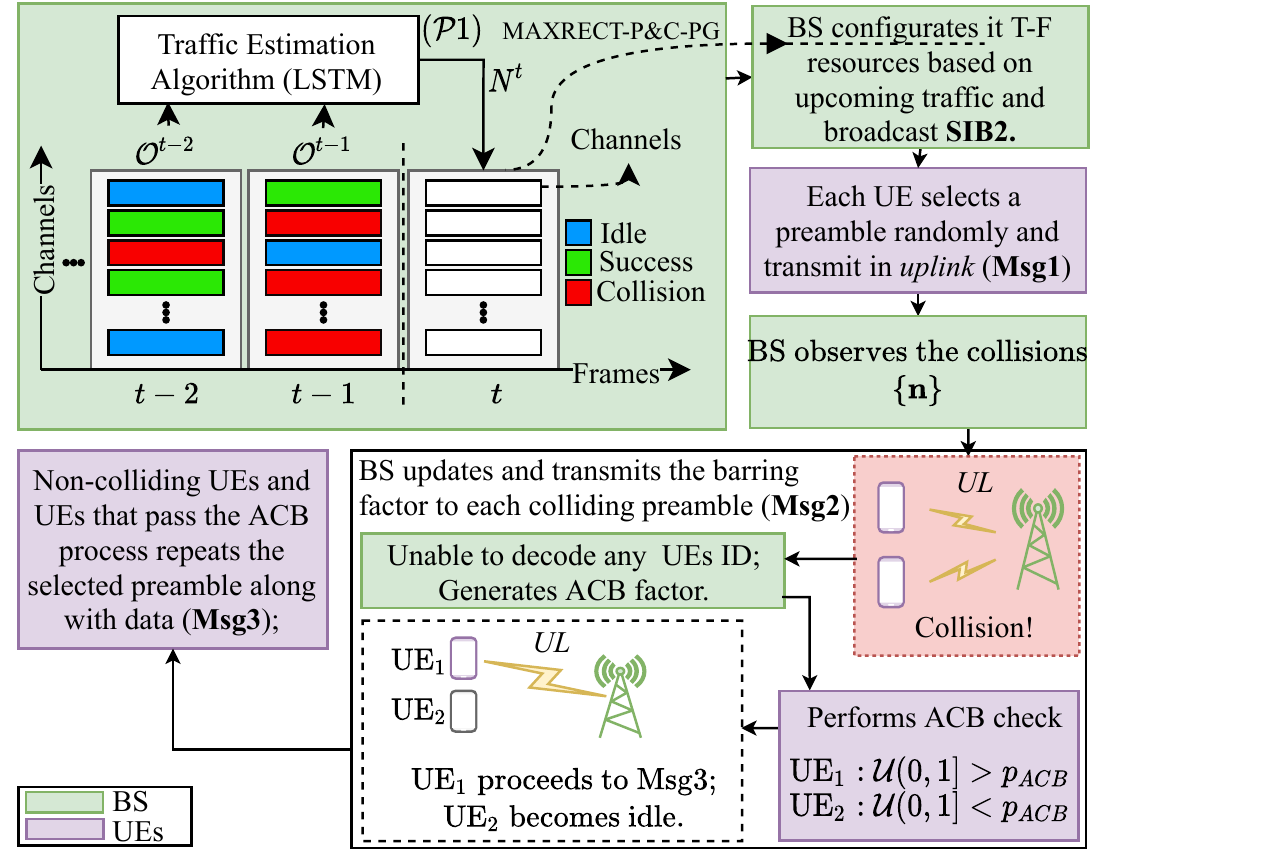}
\vspace{-3mm}
\caption{\small LSTM-ACB-based RA Scheme for hybrid traffic.} 
\label{fig:H-RL RAprot}
\vspace{-3mm}
\end{figure}

{We propose the LSTM-ACB-based RA scheme, Fig. \ref{fig:H-RL RAprot}, combining GF fast transmission and an ACB-based collision alleviation. The traffic prediction and RS procedures are performed to grant resources preferable to URLLC users and introduction of Msg2 and Msg3 alleviate the congestion. }

\begin{enumerate}[label={S.\arabic*},left=0mm,start=0]
\item  {\bf(SIB2)} System Information Broadcast -- Based on a belief state containing $\hat{K}^u$ and $\hat{K}^m$, the BS solves RA allocation $\mathcal{P}2$ to define the optimal \hls{number of} channels for each service.
\item {\bf(Msg1)} Preamble selection -- Users with data in their queue receive SIB2 and initiate the RA procedure by transmitting the preamble selection message in UL.
\item {\bf(Msg2)} Barring factor transmission -- BS observes triplet $\{V_s, V_c, V_i\}$, {broadcasts} the barring factor $p_{\rm acb}$ to each user; the colliding UEs \hls{generates} a random number and compares with $p_{\rm acb}$ to determine if they proceed to Msg3. {UEs that \hls{chose} a channel solely receives $p_{\text{acb}} = 1$.} 
\item {\bf(Msg3)} UEs that continue the RA procedure{, \it i.e.}, UEs that \hls{chose} the channels solely\hls{, and} the ones that pass the ACB {rule}, repeat the preamble and transmit their data.
\end{enumerate}

\vspace{-2mm}
\section{Results and  Discussion}\label{sec:results}
\hls{The parameters for the conducted simulations are depicted in Table I.}

\begin{table}[!htbp]
\caption{Adopted values for system and channel parameters.}
\centering
\begin{tabular}{l|l}\hline
 \bf Parameter      & \bf Value  \\ \hline
URLLC UEs   & $K^u = 25$  \\
mMTC UEs    & $K^m = 1000$  \\
{\# Frames} & {$\mathcal{F}=1200$} \\ 
Periodic mMTC UEs & $K^p_m = 10$\\
    Periodic uRLLC frames  & $T_p = 10$    \\
   \# Time RBs per frame   & $S = 10$    \\
{\#} Frequency RBs per frame & $F = {50}$\\ 
System Total Bandwidth & ${9}$ MHz\\
URLLC packet size  & $P^u = 32$ bytes  \\
mMTC packet size   & $P^m = 200$ bytes \\
{ACB factor} & {$p_{\rm acb}\in\{0.2, 0.4, 0.6, \frac{1}{\bar{n}}\}$} \\
URLLC Mod. order       & $M^u = 4$     \\
mMTC Mod. order       &  $M^m = 256$   \\
\# OFDM symbols per RB        &$\nu = 14$    \\
Weights: URLLC, mMTC, Penalty & $\{\omega^u\, \omega^m\, \omega^p\}$ = {$\{.9\,\, .05\,\, .05 \}$}  \\
Protocol Overhead & $\xi = 5$ \\
\hline
{Monte-Carlo simulation (MCS)}& {$\mathcal{T}\in[10^2; \, 10^3]$ realizations}\\
\hline 
\end{tabular}
\label{tab:SimulationParameters} \vspace{-3mm}
\end{table}


\noindent{\bf Traffic Prediction} $(\mathcal{S}1)$. {Considering $|L| = 54$ and $|L^u| = 5$, $|L^m| = 49$, we train a 20 layers LSTM through 1200 epochs and $10^4$ traffic samples composed by $\mathcal{H}^t$, and the desired outputs as exact $\Breve{K}^m,\Breve{K}^u$, as in Section \ref{subsec:BacklogPred}. \hls{The mean-squared error} over $\mathcal{T}=10^3$ \hls{is} in the range of ${\textsc{mse}^m, \textsc{mse}^u}= \{10^{-6}, 10^{-5}\}$ for {{\it low} traffic} $K^m,K^u = \{500,13\}$ and ${\textsc{mse}^m, \textsc{mse}^u}= \{10^{-3}, 10^{-2}\}$, for {{\it high} traffic}  $\{K^m,K^u\} = \{2500,63\}$. Hence, the mixed mMTC-URLLC traffic prediction is feasible with this method. }

\noindent{\bf Resource Slicing} procedure $(\mathcal{S}2)$.  In the absence of RS, the BS must structure the channels to serve any user 
{In such situation, the BS can accommodate up to 30 $16f\times 1s$ channels in T-F grid. Disregarding optimization in power gains, the heuristic \textsc{maxrect} can \textit{pack} more channels, specially as $\Breve{K}^u$ increases since $\iota ^u < \iota^m$ and different formats for $L^m$, impacting directly the \hls{number of} served users and long-term backlog.}

\begin{figure}[htbp!]
\vspace{-3mm}
\centering
\includegraphics[width=1\linewidth]{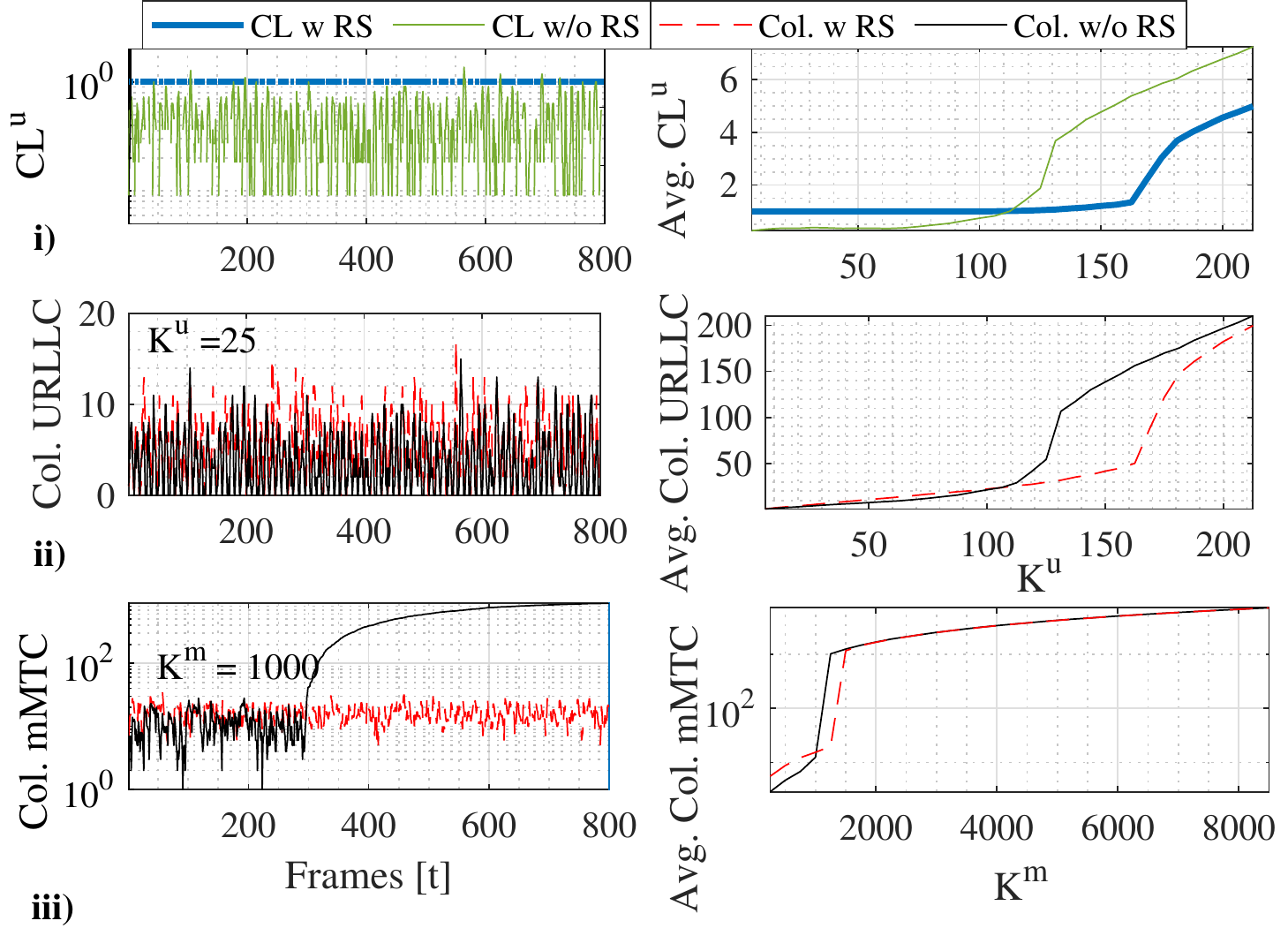}
\vspace{-6mm}
\caption{\small Performance of channel loading and collisions with RS (blue, red lines) and without RS (green, black lines) for URLLC in {\it i}) and {\it ii}), and mMTC in {\it iii}) for different \# users.}
\label{fig:ChannelLoading}
\vspace{-2mm}
\end{figure}

{Consider the \textit{channel loading} (CL) as the ratio of \textit{active users} by \textit{number of channels}, }
  $  \text{CL}^i = \frac{\Breve{K}^i}{|L^i|}, \,\, \ i \in {u,m},$ 
{depicted in Figure \ref{fig:ChannelLoading}. The $|L|$ under RS scheme with \textsc{maxrect} varies from 31 ($|L^u|=1$) to 41 ($|L^u|=40$), improving RB usage.}
Also, the CL performance using RS is capable of dedicating channels based on the upcoming traffic and \hls{optimizes} the resource utilization: 
line CL=$1$ in the subplots of Fig. \ref{fig:ChannelLoading} {\it i}) {\hls{the} occurrence of collisions under RS and fixed T-F grid is shown in {\it ii)} and {\it iii)}, which were significantly reduced in the RS scheme in priority URLLC use mode. Also, subplot \textit{iii)} presents that under a high loading ($K^m=1000$) the system is still running, although congested, with RS. }
As the loading in the system increases, the congestion in priority URLLC use mode is mitigated.%

\vspace{-2mm}
\noindent{\bf ACB-based Congestion management} ($\mathcal{P}3$) {is analysed applying LSTM-ACB-based RA protocol assuming classical $|L|=54$ channels, varying \hls{number of} users, \hls{number of} reserved URLLC channels, and applying Msg2 and Msg3 proposed scheme. The performance is averaged over $\mathcal{T}=100$ MCS and over $\mathcal{F}=10^3$ frames. We adopt {\it normalized throughput} as evaluation metric, }
$\eta=\frac{V_s}{{L}}$ where $p_{\rm acb} \in \{0.2, 0.4, 0.6, 1, \frac{1}{\bar{n}^{-1}}\}$, {$p_{\rm acb} = 1$ \hls{represents} the absence of Msg2 and Msg3, \textit{i.e.} G-F protocol, while the \textit{optimal} ACB \hls{from} Eq. \eqref{eq:opt_acb}. We vary $K^m$ from $10^3$ to $3\cdot10^4$, and $K^u$ follows the rule of proportionality.  }
Also, we consider $T_{\rm {\textsc{acb}}} = 0$, \textit{i.e.} a colliding user will retry the RA procedure in next frame. We test the {\it channels reservation} $L^u \in \{4, 5, \dots, 33, 34\}$ as $K$ increases.

\begin{figure}[!htbp]
\centering
\includegraphics[width=.384\textwidth]{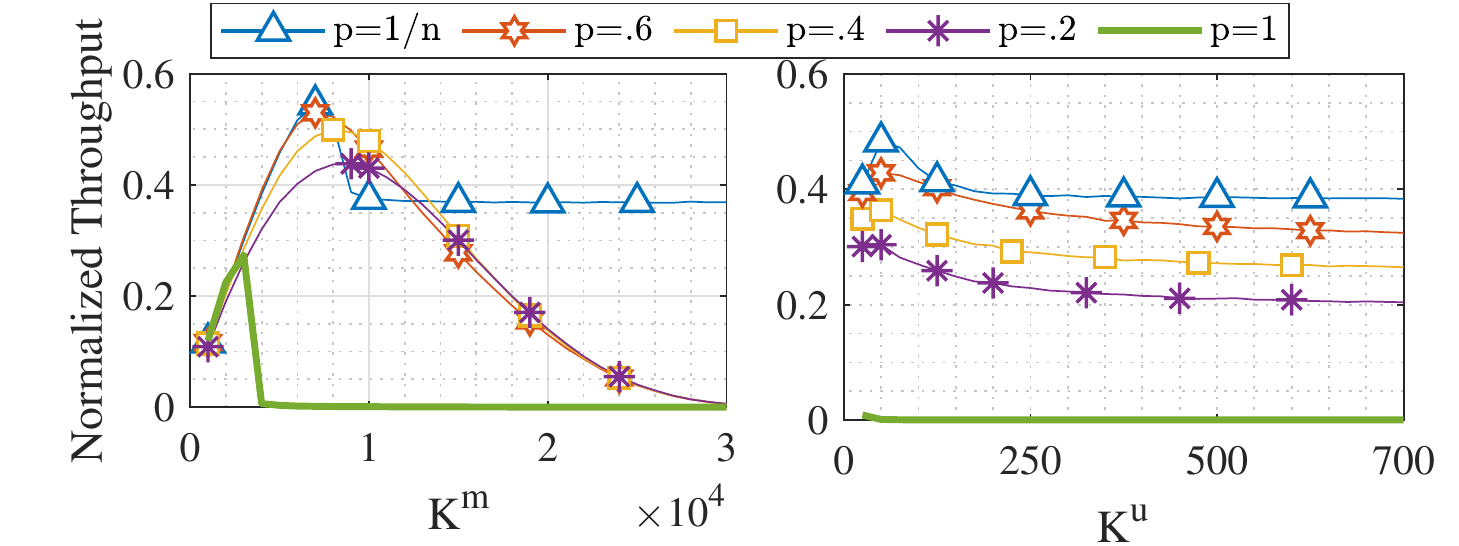}
\caption{\small Normalized Throughput $\eta$ for ACB-based contention resolution with $L = 54${, and $K^u = \frac{K^m}{{40}}$ and {$L^u \in \{4, 5, \dots, 33, 34\}$ as $K$ increases}.}}%
\label{fig:HGP_Var}
\vspace{-7mm}
\end{figure}

{Fig. \ref{fig:HGP_Var} shows $\eta$ \textit{versus} $K$ in the given scenario. Hence, in GF benchmark operating under $\{K^m \geq 4000, K^u \geq 100\}$ the system \hls{has} high congestion and the probability of access from both use modes approximate to zero due to \textit{unsolved collisions}, increasing backlog and $\Breve{K}^i \rightarrow K^i$, $i \in \{u,m\}$ as $t$ \hls{increases}. The proposed protocol, Fig. \ref{fig:H-RL RAprot} (Msg2 and Msg3), can alleviate the congestion, and the performance is increased significantly.   }
{Adopting $p^*$, the system can maintain the throughput for high \hls{number of} of users. }

\begin{figure}[!htbp]
\vspace{-2mm}
\centering
\begin{subfigure}[!htbp]{0.48\textwidth}
\centering
\includegraphics[width=.75\textwidth]{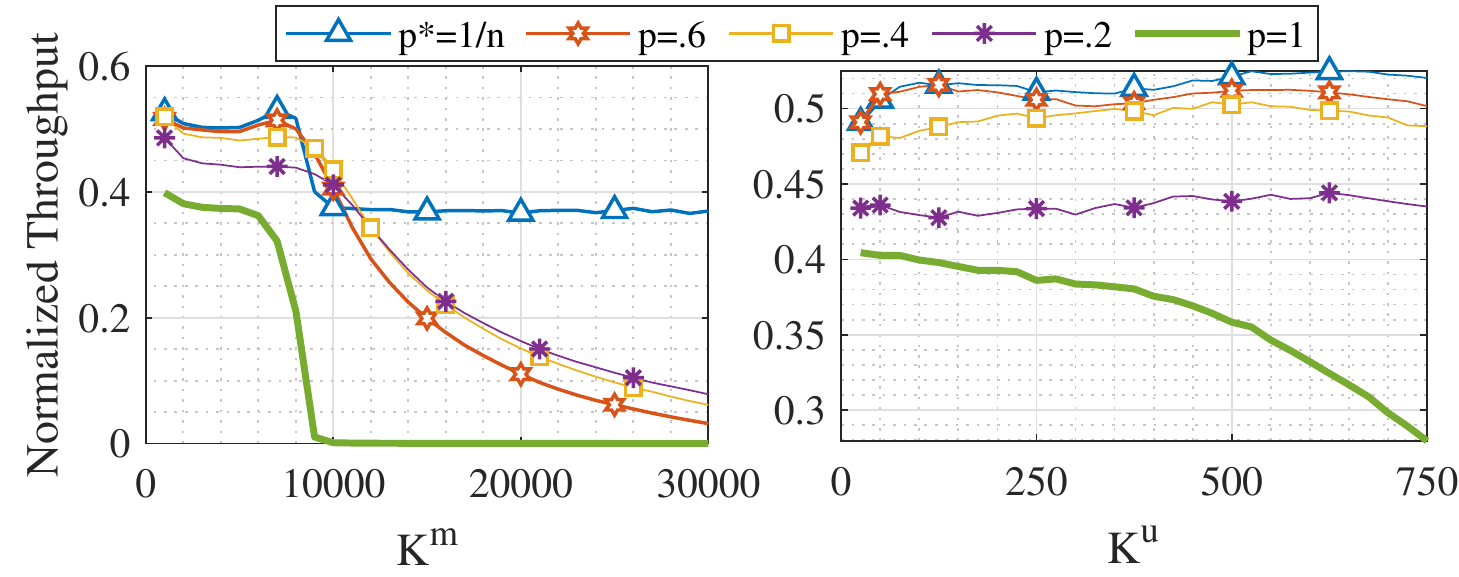}
\vspace{-2mm}
\caption{Normalized throughput.}
\label{fig:HGP_ideal_throughput}
\end{subfigure}
\hfill
\begin{subfigure}[!htbp]{0.48\textwidth}
\centering
\includegraphics[width=.75\textwidth]{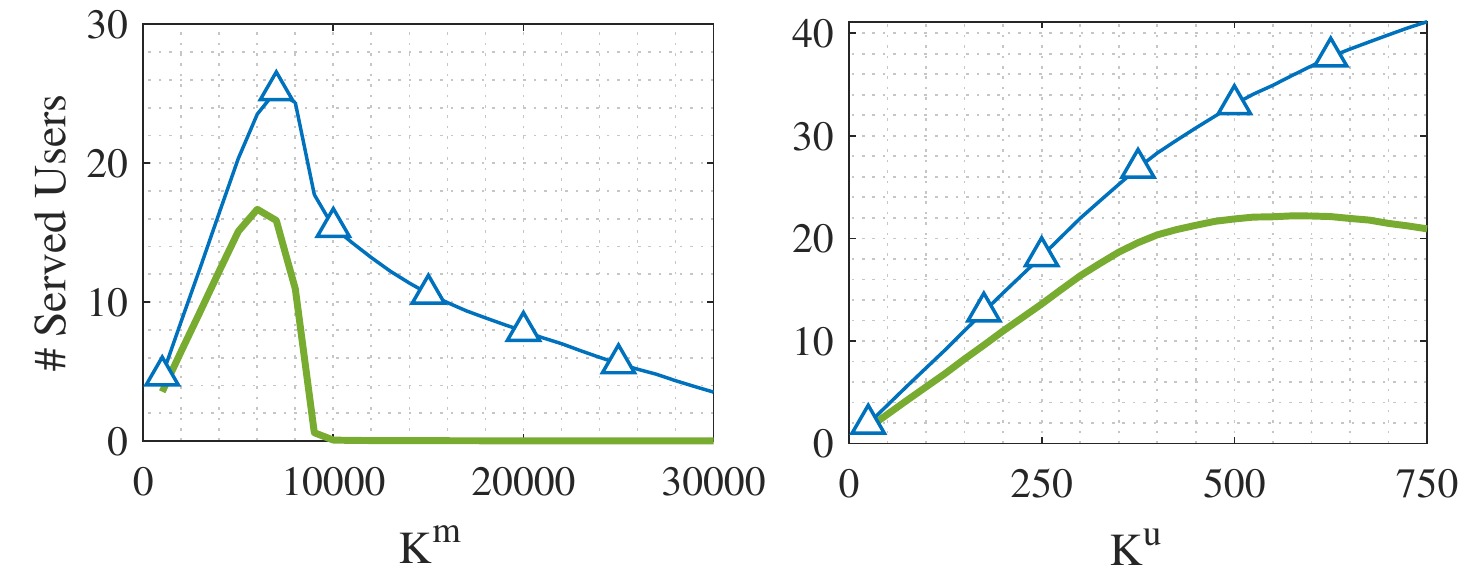}
\vspace{-2mm}
\caption{Mean number of served users.} \label{fig:HGP_ideal_servedUEs}
\end{subfigure}
\caption{\small Perfect traffic prediction and RS for both URLLC and {crowded} mMTC {use modes. $\mathcal{F}=1200$ frames}; {$K^u = \frac{K^m}{400}$}}
\label{fig:PTraffic_PAll}\vspace{-2mm}
\end{figure}

{Assuming perfect traffic prediction, perfect RS, and proposed protocol with $p^*$, Fig. \ref{fig:PTraffic_PAll} presents $\eta$ \textit{versus} $K$ for the full scheme. The $\eta_{\textsc{urlc}}$ is increased when the loading increases due to:  {\textit{i}})  less channel usage by the BS under low traffic, then the denominator of $\eta$ is  {reduced}, {\textit{ii}}) as $\Breve{K}^u$ increases, $L^u$ increases proportionally. Additionally, when $\Breve{K}^u$ is high, {the adopted policy \hls{reserves} most of the resources to this use mode, which implies \hls{in} few access opportunities} to mMTC users, causing a high collision probability and abruptly decreasing the number of served users, as indicated in Fig. \ref{fig:HGP_ideal_servedUEs} for $\{K^m,K^u\} \geq \{10000,250\}$. }

\vspace{-5mm}
\bibliographystyle{IEEEtran}
\bibliography{refs.bib}
\end{document}